\documentclass[preprint,amsmath,amssymb,aps,showpacs,showkeys]{revtex4-1}
\usepackage{graphicx}
\usepackage{dcolumn}
\usepackage{bm}
\usepackage{subeqnarray}
\usepackage{graphics}
\usepackage{amsmath}
\begin{document}
\title{Dirac states in armchair- and zigzag-edged graphene M\"{o}bius strips}

\author{J. F. O. de Souza}\email{jfernando@fisica.ufpb.br,}
\author{Claudio Furtado}\email{furtado@fisica.ufpb.br}

\affiliation{Departamento de F\'{i}sica, CCEN, Universidade Federal da Para\'{i}ba, Cidade Universit\'{a}ria, 58051-970 Jo\~ao Pessoa, PB, Brazil} 

\begin{abstract}
Edge structure plays an essential role in the nature of electronic states in graphene nanoribbons. By focusing on the interplay between this feature and non-trivial topology in the domain of the Dirac confinement problem, this paper proposes to examine how effects associated with edge shape manifest themselves in conjunction with the topological signature typical of M\"{o}bius strips within a low-energy regime. Aiming to provide an alternative to prevailing tight-binding approaches, zigzag and armchair M\"{o}bius strips are modeled by proposing compatible sets of boundary conditions, prescribing profiles of terminations in both transverse and longitudinal directions which are demonstrated to be coherent in describing consistently transverse edge patterns in combination with a proper M\"{o}bius periodicity. Of particular importance is the absence of constraints on the solution, in contrast with infinite mass analogues, as well as an energy spectrum with a characteristic dual structure responding exclusively to the parity associated with the transverse quantum number. Zigzag ribbons are predicted to possess an intrinsic mechanism for parity inversion, while the armchair ones carry the possibility of a coexistent gapless and gapped band structure. We also inspect the influence of the edge structure on persistent currents. In zigzag-edged configurations they are found to be sensitive to a length-dependent term which behaves as an effective flux. Armchair rings show a quite distinctive property: alternation of constant and flux-dependent currents according to the width of the ring, for a fixed transverse quantum number. In the flux-free case the effects of topology are found to be entirely suppressed, and conventional odd and even currents become undistinguishable.
\end{abstract}
\maketitle

\section{Introduction}

The fabrication of graphene in 2004 by Novoselov et al. \cite {Novoselov} represented the rise of a new and promising field of research in nanoscale physics. Peculiar aspects of this intriguing allotrope of carbon have demonstrated to be quite relevant in both theoretical and experimental realizations in the domain of the physics of canonical nanoscopic structures - such as quantum dots and rings, ribbons etc., indicating an enormous potential for applications and expectations towards a new generation of graphene-based electronic devices. 

A series of important findings on the electronic behavior of typical nanostructures made from graphene have been reported in the literature. Width-dependent band gap opening by electronic confinement \cite{Son,Han,Ritter}, existence of edge states in zigzag-edged configurations and width-sensitive dual electronic structure in armchair ones \cite{Fujita,Nakada,Brey,Brey2,Ritter}, valley degeneracy breaking in infinite mass Aharonov-Bohm nanorings \cite{Recher}, etc. are well-known examples of properties of great theoretical and practical interest which illustrate sufficiently well the relevance of such systems for condensed matter physics. Recently,  theoretical studies have  been employed by several authors to study the properties
of a quasiparticle confined in nanostructures,
{\it i.  e.} quantum dots \cite{silver,schn} and quantum rings\cite{petrov,gruji,costa}.   

In particular, nanorings exhibiting M\"{o}bius topology have been object of a variety of theoretical studies recently, and some noticeable results have come out. Zigzag graphene M\"{o}bius strips were predicted to behave as topological insulators, preserving edge states at zero energy under perturbation of uniform electric fields \cite{Guo}. Also, as found originally in M\"{o}bius ladders \cite{Zhao}, destructive interference suppressing partly the transmission was also demonstrated to occur in graphene \cite{Guo}. Edge magnetism was shown to play an important role in nanoribbons under such topological influence \cite{Jiang}. At the same time, general structural, optical and electronic properties of such systems have been studied extensively by several computational simulation methods \cite{Caetano,Caetano2,Wang,Korhonen}. Low-energy treatments, on the other hand, have been shown to be problematic in part. Infinite mass confinement and M\"{o}bius periodicity were proven to be compatible only by restricting the domain of the solutions, which is also manifested in physical quantities such as energies and persistent currents \cite{Souza,Souza2}. There is no low-energy approach considering edge effects in the background of M\"{o}bius topology, which is expected to be a serious candidate for a scenario where such constraints do not take place.   

In this paper, we analyze the effects of edge geometry on the electronic properties of M\"{o}bius-type graphene nanorings in the low-energy approximation. It is organized as follows. In Section \ref{sec2}, we briefly present a preliminary review on important aspects of graphene, introducing basic objects and notation utilized. In Section \ref{sec3}, we propose a model which combines transverse boundary conditions characteristic of zigzag nanoribbons and longitudinal ones capable of introducing M\"{o}bius topological character in a compatible way. General properties of bulk and edge states are examined in light of such prescription. In a similar way, an approach to the armchair case is presented in Section \ref{sec4}. Section \ref{sec5} focus on the profile of persistent currents in both configurations, with emphasis in their edge signatures. 

\section{Dirac states around $\vec{K}$ and $\vec{K}'$ points}\label{sec2}

Graphene is a two-dimensional lattice composed of carbon atoms arranged in a honeycomb structure. In spite of consisting basically of a monolayer of graphite, graphene shows very particular and important distinctions from its three-dimensional analogue from the point of view of electronic behavior. Maybe one of the most striking aspects is the emergence of a Dirac-type regime under certain conditions. Precisely, the usual tight-binding aproximation reveals that near certain points in momentum space its band structure acquires a typical conical shape, where electrons exhibit a relativistic signature and their dynamics obey a particular Dirac equation. Such points are usually known as Dirac points, and here are denoted by $\vec{K}$ and $\vec{K'}$. A particular choice of lattice orientation enables us to place them at $\vec{K}=\left(\frac{4\pi}{3\sqrt{3}a_0},0\right)$ and $\vec{K'}=\left(-\frac{4\pi}{3\sqrt{3}a_0},0\right)$, where $a_0$ denotes the typical distance between carbon atoms in the lattice. In the vicinity of $\vec{K}$, electrons must obey an effective Dirac equation written as 
\begin{equation}\label{eqDirac1}
v_F\left(\vec{\sigma}\cdot \vec{q}\right)\psi=E\psi,
\end{equation} 
where $\vec{\sigma}=\left(\sigma_x,\sigma_y\right)$, Pauli matrices, and $\vec{q}$ is the momentum relative to $\vec{K}$. On the other hand, an analogous equation is found around the $\vec{K'}$ point, differing from the first one only by a sign in $\sigma_x$. It can be written as
\begin{equation}
v_F\left(\vec{\sigma}'\cdot \vec{q}\right)\psi'=E\psi'.
\end{equation}
In this case, the matrices read $\vec{\sigma}=\left(-\sigma_x,\sigma_y\right)$, and the spinor $\psi'$ corresponds to the wave function with respect to $K'$ valley. 

In that description, spinor components are related to graphene sublattices, in such a manner that there must be correspondence between the profile of boundary conditions to be adopted and the type of sublattice at the edges to be isolated or identified. Nevertheless, in this context, we should deal with complete wave functions for the respective sublattices (see Ref. \cite{Castroneto}), namely:

\begin{equation}\label{eq3}
\Psi_{\mathcal{A}}(\vec{r})=e^{i\vec{K}\cdot \vec{r}}\psi_{\mathcal{A}}(\vec{r})+e^{i\vec{K'}\cdot \vec{r}}\psi'_{\mathcal{A}}(\vec{r})
\end{equation}
and
\begin{equation}\label{eq4}
\Psi_{\mathcal{B}}(\vec{r})=e^{i\vec{K}\cdot \vec{r}}\psi_{\mathcal{B}}(\vec{r})+e^{i\vec{K'}\cdot \vec{r}}\psi'_{\mathcal{B}}(\vec{r}),
\end{equation}
where $\psi_{\mathcal{A}(\mathcal{B})}$ and $\psi'_{\mathcal{A}(\mathcal{B})}$ represent the spinor components associated with the valleys $K$ and $K'$, respectively. 

\begin{figure}[t]
\centering
\label{fig1}
\includegraphics[scale=0.5]{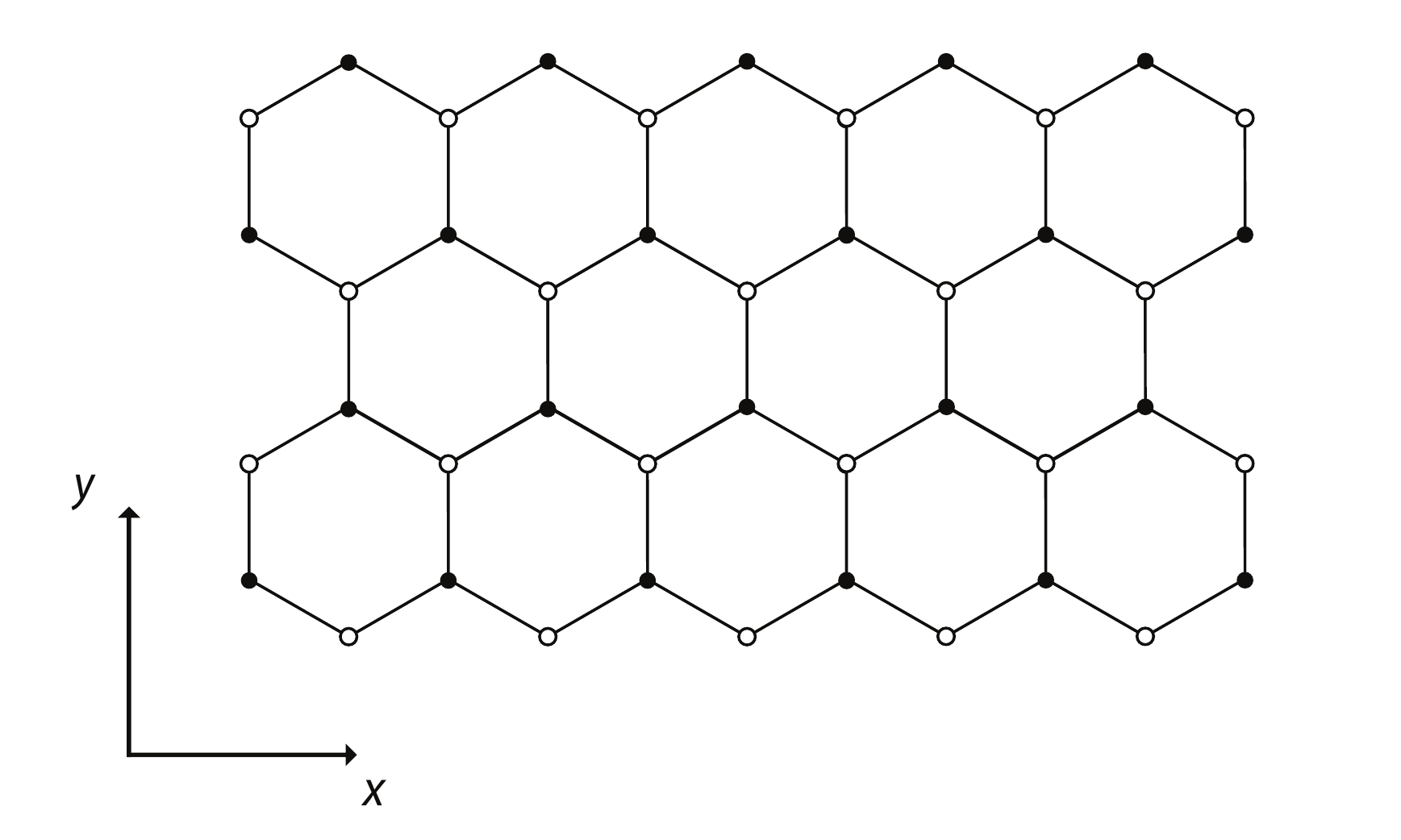}
\caption{Graphene sheet with zigzag and armchair terminations in the $y$ and $x$ directions, respectively.}
\end{figure}

\section{M\"{o}bius strips with zigzag edges}\label{sec3}

The modeling of graphene M\"{o}bius strips in the low-energy regime consists of mimicking their topological character by applying a compatible set of transverse and longitudinal boundary conditions \cite{Souza}. Within this scheme, they have been described from typical nanoribbons by neglecting the edge geometry of their transverse terminations. From now on, we intend to pay more attention to this aspect proposing analytical prescriptions for treating both zigzag- and armchair edged M\"{o}bius configurations.

As starting point, let us focus on the characterization of a zigzag M\"{o}bius-type configuration. Consider a ribbon oriented as illustrated in Fig. 1. We impose the following set of boundary conditions along the $y$ direction:
\begin{equation}
\psi_{\mathcal{A}}(x,0)=\psi_{\mathcal{B}}(x,d)=0.
\end{equation}
These conditions were first presented in Refs. \cite{Brey,Brey2} in describing the confinement of Dirac fermions in zigzag nanoribbons. After employing the zigzag profile in the transverse direction, in order to incorporate a M\"{o}bius shape into the resulting system, one proposes the following set for the longitudinal direction:
\begin{equation}\label{condmob1}
\psi_{\mathcal{A}}(0,y)=e^{iKL}\psi_{\mathcal{B}}(L,d-y)
\end{equation}
and
\begin{equation}\label{condmob2}
\psi_{\mathcal{B}}(0,y)=e^{iKL}\psi_{\mathcal{A}}(L,d-y),
\end{equation} 
which prescribes a proper M\"{o}bius topological signature, conforming the characteristic twisted periodicity to the nature of the ends at the interface. They translate the type of sublattice present at the edges to be identified along an armchair line. Fig. 2 illustrates edge and interface geometry compatible with this situation. In general, they are derived by imposing conventional twisted conditions on the complete wave functions, namely
\begin{equation}\label{compf1}
\Psi_{\mathcal{A}}(0,y)=\Psi_{\mathcal{B}}(L,d-y)
\end{equation}
and
\begin{equation}\label{compf2}
\Psi_{\mathcal{B}}(0,y)=\Psi_{\mathcal{A}}(L,d-y).
\end{equation}
Similar conditions are taken in Ref. \cite{Guo} in the context of a tight-binding approach. Nevertheless, they do not need to have this general form in a continuum treatment, which enables us to make some particular choices. Thus, by imposition of (\ref{compf1}) and (\ref{compf2}), we take the choices (\ref{condmob1}) and (\ref{condmob2}) for conditions to be satisfied by our spinor components $\psi_{\mathcal{A}}$ and $\psi_{\mathcal{B}}$.  

At this point, we proceed to analyze bulk and edge solutions separately, confining ourselves to apply a set combining zigzag and M\"{o}bius conditions as above to the solution of the Dirac equation around the $\vec{K}$ point. Let us write the system obtained from that equation:
\begin{equation}
\left\{ \begin{array}{ccc} -i\left(\frac{\partial}{\partial x}-i\frac{\partial}{\partial y}\right)\psi_{\mathcal{B}}=E\psi_{\mathcal{A}} \\ \\ -i\left(\frac{\partial}{\partial x}+i\frac{\partial}{\partial y}\right)\psi_{\mathcal{A}}=E\psi_{\mathcal{B}} \end{array} \right.,
\end{equation} 
with $E^{2}=q_x^2-k_y^2$, where $k_y=iq_y$ and $v_F=1$. Zigzag ribbons are known to have two distinct solutions depending on the nature of $k_y$ \cite{Brey,Castroneto}. Namely, $k_y$ an imaginary number is associated with a special type of state, usually called edge state, whereas $k_y$ real represents conventional confined modes. Subsections below deal with this aspect in the background of a topological M\"{o}bius strip.

\begin{figure}[t]
\centering
\label{fig2}
\includegraphics[scale=0.11]{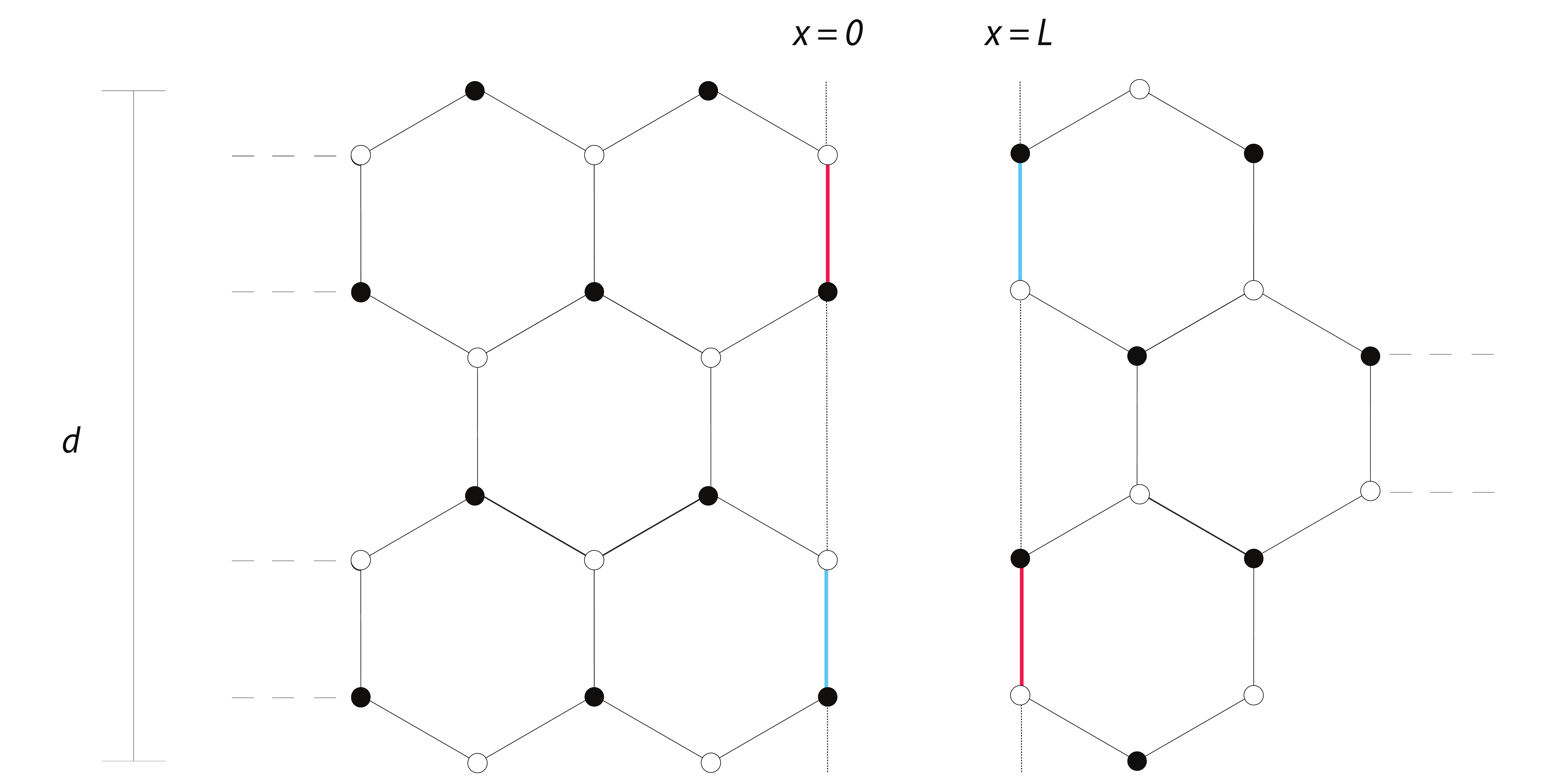}
\caption{Schematic illustration representing the identification of armchair ends located at $x=0$ and $x=L$, as prescribed by the longitudinal boundary conditions (see text).}
\end{figure}

\subsection{Bulk states}

The case with $k_y$ a complex number is characterized by typically oscillating solutions and thus describes Dirac states confined to the strip, which are what we commonly call bulk states. Therefore, imposition of zigzag conditions lead spinor components to be written in terms of ordinary trigonometric functions, as well as to the following quantization rule \cite{Brey,Castroneto}:
\begin{equation}\label{eigenbulk}
\tan\left(q_y d\right)=\frac{q_y}{q_x}.
\end{equation}
From this relation, by taking M\"{o}bius conditions as expressed in (\ref{condmob1}) and (\ref{condmob2}), we obtain
\begin{equation}
\sin\left(q_yy-\theta\right)=\sin\left(q_y(d-y)\right)e^{i\left(K+q_x\right)L}
\end{equation}
and
\begin{equation}
\sin\left(q_yy\right)=\sin\left(q_y(d-y)-\theta\right)e^{i\left(K+q_x\right)L},
\end{equation}
where we have defined $\theta=\arctan\left(\frac{q_y}{q_x}\right)$. These relations yield consistently
\begin{equation}
e^{-i\left(K+q_x\right)L}=(-1)^{\gamma},
\end{equation}
defining $\gamma=0$ when $\cos\left(q_y d\right)=-\cos\theta$ and $\gamma=1$ when $\cos\left(q_y d\right)=\cos\theta$. Such a result reinforce that the M\"{o}bius character in the low-energy approximation is manifested by means of a duality relatively to the periodicity of the spinor components. When $\gamma=0$, we deal with a periodic behavior similar to that ones found in ordinary rings, whereas antiperiodicity is defined for $\gamma=1$. It must be highlighted that, in opposition to infinite mass systems described in Refs. \cite{Souza,Souza2}, constraints on the solution are not observed here, which means no restriction on the positions along the strip. At the same time, such feature is also reflected in the energy spectrum, which reads explicitly as
\begin{equation}\label{spectz}
E_Z^2=\left\{ \begin{array}{lll} \left(\frac{2\pi}{L}m-K\right)^2+q_y^2, & \textrm{$\gamma=0$} \\ \\ \left[\frac{2\pi}{L}\left(m+\frac{1}{2}\right)-K\right]^2+q_y^2, & \textrm{$\gamma=1$} \end{array} \right..
\end{equation}
with $m$, the longitudinal quantum number, an integer and $q_y$ satisfying Eq. (\ref{eigenbulk}). As can be seen, although the constraint is absent here, it must be noticed that the index regulating the passage from periodic to antiperiodic levels remains sensitive to the transverse quantization rule. As in that case, such a particular signature may also be understood as coexistent electronic structures, as long as $\gamma$ is subject to control. Here we shall call that property of $\gamma$ parity just for analogy with previous treatments \cite{Souza,Souza2}.

It must also been noticed that the presence of the $K$ term in (\ref{spectz}) exchanges periodicity relatively to $\gamma$ depending on the length of the ribbon, working effectively as mechanism for parity inversion. In fact, since $K=\frac{4\pi}{3\sqrt{3}a_0}$ and $L$ is multiple of $a=3\sqrt{3}a_0$, it becomes possible to make specific choices for $L$ such that the longitudinal part in the spectrum alternates between periodic and antiperiodic contributions for a given expression. For instance, for $L=(2N+1)a$, $N$ integer, such contribution adds $1/2$ to both expressions in (\ref{spectz}), which thus acts permuting their parities. The case with $L=2Na$ does not fulfill that feature. 

It should be pointed out that adding an Aharonov-Bohm-type flux must be another mechanism to manipulate that dual electronic character conveniently. For example, by taking a half flux quantum, we effectively exchange parity in the spectrum, as already pointed in Ref. \cite{Souza2}.

\subsection{Edge states}

For $k_y$ a real number, we deal with solutions that correspond to edge states of the system \cite{Brey}, which are localized at the edges of the strip. Just as in the preceding subsection, the imposition of zigzag conditions leads to the following relation:
\begin{equation}
e^{2k_y d}=\frac{q_x+k_y}{q_x-k_y}.
\end{equation}
In contrast, in this case we have hyperbolic functions as solutions instead of trigonometric ones. Explicitly:
\begin{equation}
\psi_{\mathcal{A}}(x,y)\propto e^{k_y d}\sinh\left(k_y y-k_y d\right)e^{iq_x x}
\end{equation}
and
\begin{equation}
\psi_{\mathcal{B}}(x,y)\propto \left(q_x+k_y\right)\sinh\left(k_y y\right)e^{iq_x x}.
\end{equation}
Here immediately we observe they differ from the bulk solutions especially because they do not have an oscillating behavior. Indeed, they exhibit a typically hyperbolic behavior which corresponds physically to the localization of such states at the edges. Finally, taking into account this picture, in order to characterize such states as edge states of a M\"{o}bius strip, we apply the M\"{o}bius conditions to the spinor components $\psi_{\mathcal{A}}$ and $\psi_{\mathcal{B}}$ above. Both conditions yield the same relation
\begin{equation}\label{quantedge}
e^{2i(K+q_x)L}=1,
\end{equation}
which leads to the quantization rule $\left(q_x+K\right)L=m\pi$, where $m$ is an integer. Notice that the edge states are more numerous than in periodic rings, where such a relation reads $(q_x+K)L=2m\pi$. States corresponding to even multiples of $\pi$ are present in both cases; for the M\"{o}bius case, however, in addition we identify the presence of intermediate levels which can be thought as associated with an inherent antiperiodicity. At this moment, we note that these antiperiodic levels effectively exhibit periodic nature; also, an index defining and regulating such duality of periodicity is absent here, in contrast with the other cases. Another fact to be highlighted is that the relation (\ref{quantedge}) may be written as $2L$-periodic condition on the spinor, namely: $\psi(0,y)=\psi(2L,y)$. This feature appears to express explicitly a well-known property of M\"{o}bius topology. As we know, constructing a M\"{o}bius strip from a generic ribbon involves a characteristic process which consists of joining its two ends after twisting one of them by $\pi$, resulting in a peculiar geometric object having effectively only one edge with twice its original length.

\section{M\"{o}bius strips with armchair edges}\label{sec4}

Graphene nanoribbons with armchair edges are usually described by means of other set of boundary conditions \cite{Brey,Brey2}. In this case, conditions mixing $\vec{K}$ and $\vec{K'}$ valleys are employed:
\begin{equation}\label{armchair1}
\psi_{\mu}(0,y)+\psi'_{\mu}(0,y)=0
\end{equation} 
and
\begin{equation}\label{armchair2}
e^{iKL}\psi_{\mu}(L,y)+e^{-iKL}\psi'_{\mu}(L,y)=0,
\end{equation} 
with $\mu=\mathcal{A},\mathcal{B}$. Consider a nanoribbon with length $d$ and width $L$, with armchair terminations in the $x$ direction, as indicated in Fig. 3. These conditions yield the quantization rule $q_x=\frac{1}{L}n\pi -K$, where $n$ is an integer, in agreement with that in Refs. \cite{Brey,Castroneto}.

In order to construct a M\"{o}bius strip from a nanoribbon characterized by means of the conditions above, we propose a M\"{o}bius-type periodicity in the following form:
\begin{equation}
\Psi_{\mathcal{A}}(x,0)=\Psi_{\mathcal{A}}(L-x,d).
\end{equation}
Adopting only one condition without mixing the sublattices is compatible with the referred physical arrangement of the atoms, since it mimics exactly a zigzag-Klein interface in the $y$ direction, corresponding to configurations with armchair-type transverse terminations. This situation can be visualized in Fig. 2. Such condition reads as follows:
\begin{equation}
e^{iKx}\psi_{\mathcal{A}}(x,0)+e^{-iKx}\psi'_{\mathcal{A}}(x,0)=e^{iK(L-x)}\psi_{\mathcal{A}}(L-x,d)+e^{-iK(L-x)}\psi'_{\mathcal{A}}(L-x,d).
\end{equation}
By taking it for a nanoribbon obtained from the conditions (\ref{armchair1}) and (\ref{armchair2}), we arrive at the following expression for quantization rule of the longitudinal component $q_y$:
\begin{equation}
e^{-iq_y d}=(-1)^{n}.
\end{equation}
As can be seen, energy eigenvalues are sensitive to the momentum quantization in the $x$ direction, exhibiting an alternating behavior in dependence on the parity of the quantum number $n$. For $n$ odd, the longitudinal momentum reads $q_y=\frac{1}{L}\left(2m+1\right)\pi$, whereas $q_y=\frac{2}{L}m\pi$ for $n$ even, where $m$ is an integer. Thus, the energy spectrum is written as
\begin{equation}\label{spect}
E_A^2=\left\{ \begin{array}{lll} \frac{4\pi^2}{d^2}m^2+\left(\frac{n\pi}{L} -K\right)^2, & \textrm{$n$ even} \\ \\ \frac{\pi^2}{d^2}\left(2m+1\right)^2+\left(\frac{n\pi}{L} -K\right)^2, & \textrm{$n$ odd} \end{array} \right..
\end{equation}
As in the case of zigzag ribbons, notice that there is no spatial restriction on the solution; also, such twofold periodicity of the spectrum may be thought as expression of a peculiar electronic behavior which is characterized by the presence of a dual gap defined according to the parity of the transverse quantum number. Likewise, the implicit assumption here is the possibility of external control of such a feature.
\begin{figure}[t]
\centering
\label{fig2}
\includegraphics[scale=0.11]{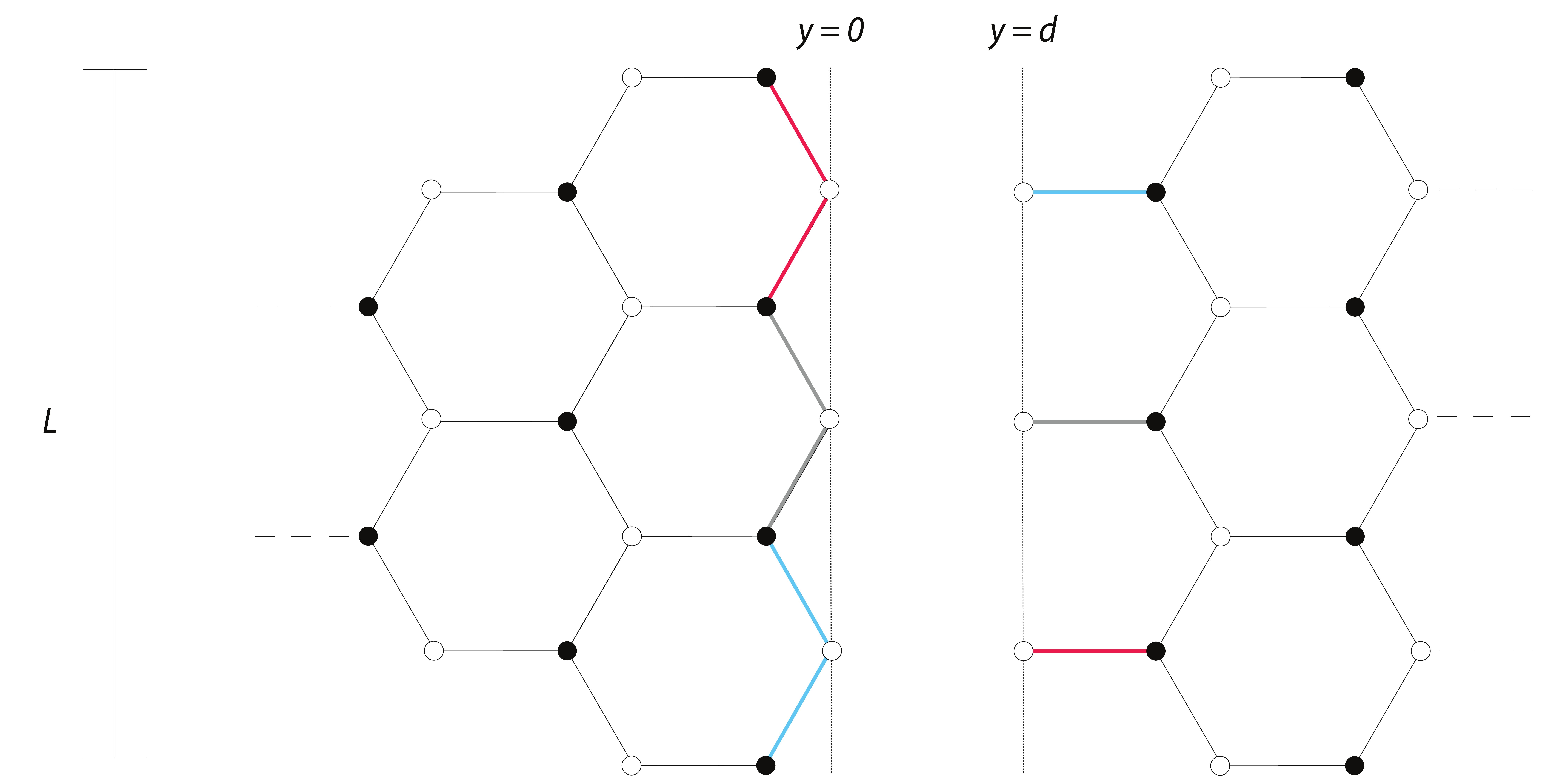}
\caption{Schematic illustration representing the identification of zigzag and Klein ends located at $y=0$ and $y=d$, respectively, as prescribed by the longitudinal boundary condition (see text).}
\end{figure}

In addition, another related feature deserves attention. Looking at the periodic part of the spectrum ($n$ even), we notice that it is possible to set up a zero gap by taking $K=\frac{n\pi}{L}$. Since $E=\frac{2\pi m}{d}$, $m=0$ guarantees zero energy and thus the bands touch. It means that the strip behaves as a conductor for specific values of width. The same result is found for nanoribbons with armchair edges \cite{Fujita,Nakada,Brey2}, differing by the fact that longitudinal momentum is not quantized in that case; a similar one is also supported by nanotubes \cite{Hamada,Saito,Kane}. On the other hand, for the antiperiodic one ($n$ odd), the longitudinal term is not cancelled, which would work effectively as a gap in case of being feasible to isolate a given parity. Therefore, the existence of mechanisms capable of manipulating conveniently such parity would give rise to a unique electronic property: metallic and semiconducting band structures whose alternance does not involve necessarily changes in size as in nanoribbons.  

\section{Edge signature in persistent currents}

The effect of M\"{o}bius topology on persistent currents is known to be manifested by a characteristic dual structure whose regulation is directly connected with the profile of transverse confinement adopted \cite{Yakubo,Souza2}. By introducing a typical Aharonov-Bohm flux \cite{Aharonov}, i.e., a thin encircled flux piercing perpendicularly at the center of the ring configuration, an expression for these entities at zero temperature can be obtained by calculating
\begin{equation}
\mathcal{I}=-\frac{\partial \mathcal{E}}{\partial \phi},
\end{equation}
where $\mathcal{E}=\sum_{n,m} E_{nm}$ is the ground state energy and $\phi$ the Aharonov-Bohm flux. 

In particular, since topology in this case acts effectively splitting the spectrum, specifically giving rise to a discrepancy in the longitudinal part of the energy eigenvalues, and the $K$ term inserts itself into the Hamiltonian in different ways according to the direction chosen, currents in zigzag and armchair structures are expected to experience quite distinct responses to changes in size. 

In the zigzag case, bulk solutions incorporate the same character found for nanoribbons with infinite mass profile, i.e., a dual structure explicitly defined. In contrast, edge states exhibit only that periodic behavior, which means that currents must be seen as in a $2L$ periodic ring. Let us consider the two cases separately. In the bulk case they are expressed as
\begin{equation}
\mathcal{I}_{Z_b}^{(0)}=\frac{4\pi^2}{L^2\phi_0}\sum_{n,m}\left[\frac{2\pi}{L}\left(m-\frac{\phi}{\phi_0}\right)-K\right]\left\{\left[\frac{2\pi}{L}\left(m-\frac{\phi}{\phi_0}\right)-K\right]^2-k_y^2\right\}^{-1/2},
\end{equation} 
for $\gamma=0$, and
\begin{equation}
\mathcal{I}_{Z_b}^{(1)}=\frac{4\pi^2}{L^2\phi_0}\sum_{n,m}\left[\frac{2\pi}{L}\left(m+\frac{1}{2}-\frac{\phi}{\phi_0}\right)-K\right]\left\{\left[\frac{2\pi}{L}\left(m+\frac{1}{2}-\frac{\phi}{\phi_0}\right)-K\right]^2-k_y^2\right\}^{-1/2}
\end{equation}
for $\gamma=1$, where $\phi_0$ is the flux quantum. As already mentioned, the $K$ term here is inserted as a longitudinal contribution to the currents, behaving as an effective Aharonov-Bohm flux. We can define a total effective flux as $\Phi=\phi+\frac{L\phi_0}{2\pi}K$, consisting of an actual interacting flux and a fictitious one. This additional term determines that $L$ can be chosen conveniently to adjust currents just as a real Aharonov-Bohm flux does. From the eigenvalues associated with edge solutions, on the other hand, we obtain:
\begin{equation}
\mathcal{I}_{Z_e}=\frac{4\pi^2}{L^2\phi_0}\sum_{n,m}\left[\frac{\pi}{L}\left(m-\frac{\phi}{\phi_0}\right)-K\right]\left\{\left[\frac{\pi}{L}\left(m-\frac{\phi}{\phi_0}\right)-K\right]^2-k_y^2\right\}^{-1/2}. 
\end{equation}

Let us now focus on the armchair case. Besides the characteristic parity sensitivity, another particularly interesting feature is a width-dependent alternance of signatures for both even and odd currents. To see it clearly consider their expressions:
\begin{equation}
\mathcal{I}_{A}^{(e)}=\frac{4\pi^2v_F}{d^2\phi_0}\sum_{n,m}\left(m-\frac{\phi}{\phi_0}\right)\left[\frac{4\pi^2}{d^2}\left(m-\frac{\phi}{\phi_0}\right)^2+\left(\frac{n\pi}{L}-K\right)^2\right]^{-1/2},
\end{equation}
for $n$ even, and
\begin{equation}
\mathcal{I}_{A}^{(o)}=\frac{4\pi^2v_F}{d^2\phi_0}\sum_{n,m}\left(m+\frac{1}{2}-\frac{\phi}{\phi_0}\right)\left[\frac{4\pi^2}{d^2}\left(m+\frac{1}{2}-\frac{\phi}{\phi_0}\right)^2+\left(\frac{n\pi}{L}-K\right)^2\right]^{-1/2},
\end{equation}
for $n$ odd. For $n$ fixed, $L=\frac{n\pi}{K}$ turns currents insensitive to the Aharonov-Bohm flux. It means that, depending on the dimensions of the strip, such quantities in armchair configurations may be controled in order to behave as constant functions. Since those particular choices for $L$ results in vanishing a term common to both expressions, there is no differentiation between odd and even currents in this case, thus eliminating the effect of topology; such a behavior is unusual for M\"{o}bius nanorings in general. Finally, we notice that not only this distinctive property but also that ones from the zigzag case are not observed in currents in graphene-based infinite mass rings \cite{Recher,Souza2}.

\section{Conclusions}\label{sec5}

In this paper, we have examined the combined influence of edge signature and topology on graphene nanostructures within the low-energy approximation. By proposing appropriate sets of boundary conditions capable of introducing both contributions simultaneously in a compatible way, we studied the properties of Dirac states for zigzag and armchair nanoribbons in response to the topological character proper to M\"{o}bius strips. 

Of particular importance is the absence of spatial constraints on the solutions, differing from their infinite mass analogues - where is necessary to reduce the domain in order to obtain a consistent picture \cite{Souza}. This characteristic reflects in the energy spectra, which shows sensitivity only to the type of transverse confinement performed.

Despite having in common the characteristic dual structure proper to the topological signature in question, Dirac states are found to exhibit appreciable differences depending on the type of transverse termination adopted. Bulk states in zigzag-edged configurations incorporate such duality explicitly, having an index responsible for regulating periodicity. Edge states do not observe the same property. States in the armchair case are similar to the bulk in the zigzag one, except for the form of that index.  

Persistent currents preserve the same structure found in the spectrum, revealing significant differences depending on the edge shape. Zigzag M\"{o}bius strips possess an extra term which can be treated as an fictitious Aharonov-Bohm flux, operating as mechanism for controling currents. The effects of topology are absent for specific values of width in armchair strips, situation in which currents exhibit a profile independent of flux.

\end{document}